\documentclass{article}

\usepackage{arxiv}

\usepackage[utf8]{inputenc} 
\usepackage[T1]{fontenc}    
\usepackage[hidelinks]{hyperref}       
\usepackage{url}            
\usepackage{booktabs}       
\usepackage{amsfonts}       
\usepackage{nicefrac}       
\usepackage{microtype}      
\usepackage{lipsum}		
\usepackage{graphicx}
\usepackage{tabularx}
\usepackage{natbib}
\usepackage{amsmath}
\usepackage{doi}
\usepackage{listings}
\lstdefinestyle{codestyle}{
    language=fortran,
    basicstyle={\ttfamily\small},
    frame={lines},
    float=[!t],
    linewidth=\linewidth,
    breaklines=true,
    breakatwhitespace=true
}
\title{Learning-Augmented Performance Model for Tensor Product Factorization in High-Order FEM}

\author{ {Xuanzhengbo Ren} \\
	Graduate School of Informatics\\
	Nagoya University\\
    Nagoya, Japan \\
	\texttt{ren@hpc.itc.nagoya-u.ac.jp} \\
	\And
	{Yuta Kawai} \\
    RIKEN R-CCS\\
	Kobe, Japan\\
    \And
    {Tetsuya Hoshino} \\
    Information Technology Center\\
    Nagoya University \\
    Nagoya, Japan \\
    \And
    {Hirofumi Tomita} \\
    RIKEN R-CCS\\
	Kobe, Japan\\
    \And
    {Takahiro Katagiri} \\
    Information Technology Center\\
    Nagoya University \\
    Nagoya, Japan \\
    \And
    {Daichi Mukunoki} \\
    Information Technology Center\\
    Nagoya University \\
    Nagoya, Japan \\
    \And
    {Seiya Nishizawa} \\
    RIKEN R-CCS\\
	Kobe, Japan\\
\thanks{This work has been submitted to the IEEE for possible publication. Copyright may be transferred without notice, after which this version may no longer be accessible.}}

\renewcommand{\shorttitle}{Learning-Augment Performance Model for Tensor Product Factorization in High-Order FEM}

\hypersetup{
pdftitle={Learning-Augmented Performance Model for Tensor Product Factorization in High-Order FEM},
pdfsubject={cs.DC},
pdfauthor={Ren, Kawai, et al.},
pdfkeywords={High-Order FEM, Tensor Product Factorization, Performance Modeling, Instruction-Level Parallelism, Learning-Augmented Modeling},
}

\begin{document}
\maketitle

\begin{abstract}
Accurate performance prediction is essential for optimizing scientific applications on modern high-performance computing (HPC) architectures. Widely used performance models primarily focus on cache and memory bandwidth, which is suitable for many memory-bound workloads. However, it is unsuitable for highly arithmetic intensive cases such as the sum-factorization with tensor $n$-mode product kernels, which are an optimization technique for high-order finite element methods (FEM).
On processors with relatively high single instruction multiple data (SIMD) instruction latency, such as the Fujitsu A64FX, the performance of these kernels is strongly influenced by loop-body splitting strategies. Memory-bandwidth-oriented models are therefore not appropriate for evaluating these splitting configurations, and a model that directly reflects instruction-level efficiency is required. To address this need, we develop a dependency-chain–based analytical formulation that links loop-splitting configurations to instruction dependencies in the tensor $n$-mode product kernel. We further use XGBoost to estimate key parameters in the analytical model that are difficult to model explicitly. Evaluations show that the learning-augmented model outperforms the widely used standard Roofline and Execution-Cache-Memory (ECM) models. On the Fujitsu A64FX processor, the learning-augmented model achieves mean absolute percentage errors (MAPE) between 1\% and 24\% for polynomial orders ($P$) from 1 to 15. In comparison, the standard Roofline and ECM models yield errors of 42\%–256\% and 5\%–117\%, respectively. On the Intel Xeon Gold 6230 processor, the learning-augmented model achieves MAPE values from 1\% to 13\% for $P=1$ to $P=14$, and 24\% at $P=15$. In contrast, the standard Roofline and ECM models produce errors of 1\%–73\% and 8\%–112\% for $P=1$ to $P=15$, respectively.
\end{abstract}

\keywords{High-Order FEM, Tensor Product Factorization, Performance Modeling, Instruction-Level Parallelism, Learning-Augmented Modeling}

\section{Introduction}\label{sec:introduction}
Performance modeling is an effective technique for gaining insights into program behavior. Developers of scientific applications can use performance models to pinpoint bottlenecks and evaluate the effectiveness of optimization strategies. On modern computer systems, the floating-point capability of CPUs is often much higher than the data transfer capability of memory. As a result, many kernels, such as Sparse Matrix-Vector Multiplication (SpMV), stencil computations, and vector triad operations along with a wide range of scientific applications, typically encounter memory-bound performance issues. Consequently, developers often rely on performance models such as Roofline \cite{williams_roofline_2009} and Execution-Cache-Memory (ECM) \cite{hofmann_execution-cache-memory_2015}, which mainly focus on memory and cache bandwidth.

With advances in semiconductor technology, however, processor designs are increasingly narrowing the gap between floating-point throughput and memory bandwidth. For example, consumer CPUs such as the AMD Ryzen 9800X3D integrate large caches, while high performance computing (HPC) processors like the Fujitsu A64FX and Intel Sapphire Rapids employ high-bandwidth memory (HBM). These architectural features reduce the dominance of memory bandwidth, making other design factors, such as pipeline depth, instruction latency, and branch prediction, more influential for application performance. Taking the Fujitsu A64FX as an example: although it is equipped with HBM, its relatively high single instruction multiple data (SIMD) instruction latency (9 cycles for a fused multiply-add operation), together with its small reservation station and limited register file size, constrain its out-of-order (OoO) execution capability. In such cases, application performance is not only limited by memory or cache bandwidth but also by in-core efficiency. Meanwhile, modern workloads are also increasingly arithmetic-intensive. For example, deep learning networks involve a large number of tensor operations that are dominated by arithmetic computation. As a result, in-core execution efficiency has become as important as memory and cache performance.

In scientific software applications, the finite element method (FEM) is one of discretization methods used to solve partial differential equations. The discretization accuracy can be improved by increasing the polynomial order ($P$) within each element. For multidimensional problems, sum-factorization enables a significant reduction in computational cost when the polynomial expansions posses a tensor-product structure (e.g., in hexahedral elements). This technique is widely regarded as a standard optimziation strategy in high-order FEM  \cite{bressan_sum_2019, badger_sum_2020}. The core computational kernel of sum factorization is a tensor $n$-mode product that contains a long inner product within its loop body, and therefore directly encounters the limited OoO execution efficiency issue as described above. 
Prior work \cite{ren_performance_2025} demonstrated that performance can be improved by splitting this inner product into multiple parts. However, because the length of the inner product varies with $P$, it is difficult to define a unified splitting strategy. As a result, a trial-and-error approach was used to determine the optimal number of splits and the length of each segment for different polynomial orders. The study suggested that performance degradation arises from data dependencies in the loop body and that splitting alleviates this issue. However a detailed explanation of why splitting improves performance remains missing. Additionally, the trial-and-error process is inefficient for practical tuning, motivating the need for a more systematic and predictive approach.

Developing such a predictive approach faces two main challenges. First, general memory-bandwidth-oriented models such as the standard Roofline model cannot fully explain the performance of our tensor kernel. For example, in the cases studied in \cite{ren_performance_2025}, the Bytes per FLOP (B/F) ratios of the tensor kernels for $P=7$ to $P=11$ range from 0.7 to 0.5, indicating that the kernel is arithmetic-intensive. On the Fujitsu A64FX processor, whose B/F ratio for L1 cache is 4, the workload’s B/F is much smaller than the machine’s B/F, implying that the kernel should be compute-bound. Yet, despite this prediction, performance still varies significantly with different loop-body splitting strategies, revealing that bandwidth-oriented models alone cannot capture the instruction-level behavior that governs performance in practice. Second, current analytical models do not incorporate loop-body splitting configurations, such as the number of splits and the length of each segment. These are essential for understanding and predicting the performance changes observed in practice.

Although several frameworks and models for tuning tensor operations vary widely in scope and methodology, they are not directly applicable to our setting. For example, Fang et al. \cite{fang_swdnn_2017} developed a performance model for tensor operations in deep learning workloads on a specific processor. Although their work mentions instruction-level optimizations, the model primarily emphasizes data transfers across memory, caches, and registers. TVM \cite{chen_tvm_2018}, a machine learning-based compilation framework, offers AutoTVM \cite{chen_learning_2018}, an automated tuning system that generates high-performance tensor kernels using statistical cost models with hardware awareness. However, AutoTVM is designed for deep learning workloads and is difficult to apply directly to sum factorization implementations in high-order FEM. Swirydowicz et al. \cite{swirydowicz_acceleration_2019} proposed a performance model for accelerating tensor products in high-order finite element methods, but their Roofline-like formulation does not capture instruction-level effects. Moreover, commonly discussed tensor-operation optimizations focus on loop tiling or loop reordering. By contrast, loop-body splitting, which is the topic examined in this research, is rarely discussed in the literature.

To address these issues, we first develop a dependency-chain–based analytical model to interpret how loop-body splitting affects performance, particularly on the Fujitsu A64FX processor. We then collect performance data from both the Fujitsu A64FX and Intel Xeon Gold 6230 processors. To enable the realistic performance prediction, the analytical model with XGBoost \cite{chen_xgboost_2016}, a decision-tree–based machine learning approach, to enable realistic performance prediction. Finally, we compare our results with predictions obtained from the Roofline and ECM models.

The contributions of this work are summarized as follows:
\begin{itemize}
\item We provide a detailed explanation of loop-body splitting, a less-explored optimization technique that can enhance OoO execution efficiency.
\item We propose an analytical model to interpret the performance variation caused by loop-body splitting in tensor $n$-mode product kernels from the perspective of dependency chains. The analytical model is further extended through a learning-augmented approach that incorporates both machine and code features. 
\item We demonstrate a promising approach for modeling performance of arithmetic-intensive kernels in scientific computing, which may serve as a useful reference for performance analysis in high-order FEM.
\end{itemize}

The remainder of this paper is organized as follows. Section II provides the background and motivation for this work. We explain sum factorization and loop-body splitting. Section III introduces the dependency-chain-based model and verifies it using microbenchmarks. Section IV describes the learning-augmented model that builds upon the dependency-chain formulation. Section V evaluates the feasibility and predictive accuracy of the learning-augmented model and compares it with the widely used Roofline and ECM models. Section VI discusses related work. Finally, Section VII concludes the paper and outlines future directions.

\section{Background and Motivation}\label{sec:background}
This study uses SCALE-DG \cite{kawai_numerical_2023, kawai_development_2025}, an atmospheric dynamical core based on the discontinuous Galerkin method (DGM). The DGM is a class of finite element methods. In this section, we introduce how the simulation programs are accelerated through the sum-factorization technique and how additional speedup is achieved by splitting the loop body in the tensor $n$-mode product kernel. This section provides the background and motivation for the present study.

\subsection{Accelerating SCALE-DG via Sum Factorization}
The DGM is attractive for HPC because its element-wise formulation localizes most computations within each element, with communication occurring only through fluxes across element interfaces. This property makes the method naturally parallelizable and well suited to distributed-memory architectures. The element-wise independence also reduces the need for global solves, which is a major advantage over continuous finite element methods.
 
\begin{figure}[t!]
  \centering
  \includegraphics[width=0.6\linewidth]{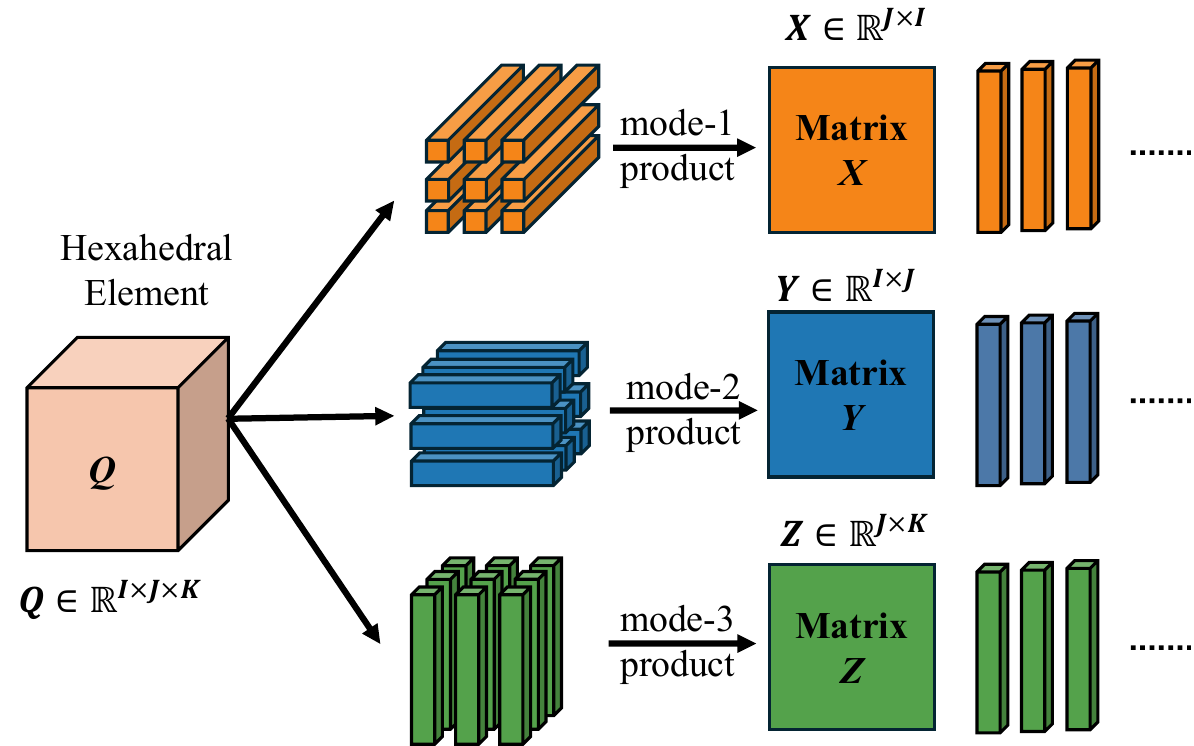}
  \caption{Illustration of sum factorization on hexahedral element.}\label{fig:demosumfactrization}
\end{figure}

However, the matrix-vector multiplications appear when evaluating the volume and surface terms and applying modal filters. In the general implementations for arbitrary elements, the size of operator matrix is $O((P+1)^6)$, where $P$ is the polynomial order. As $P$ increases, it become extremely large, resulting in substantial computational cost and memory pressure per element. For hexahedral elements, sum factorization is commonly applied to alleviate these issues by exploiting the tensor-product structure of the basis functions. As shown in Fig.~\ref{fig:demosumfactrization}, sum factorization unfolds the element along each spatial direction and performs a mode-$n$ product with a corresponding one-dimensional operator matrix. This reduces the storage requirement from $O((P+1)^6)$ to $O(3 \times (P+1)^2)$, while simultaneously lowering the computational cost.

\begin{figure}[t!]
  \centering
  \includegraphics[width=0.4\linewidth]{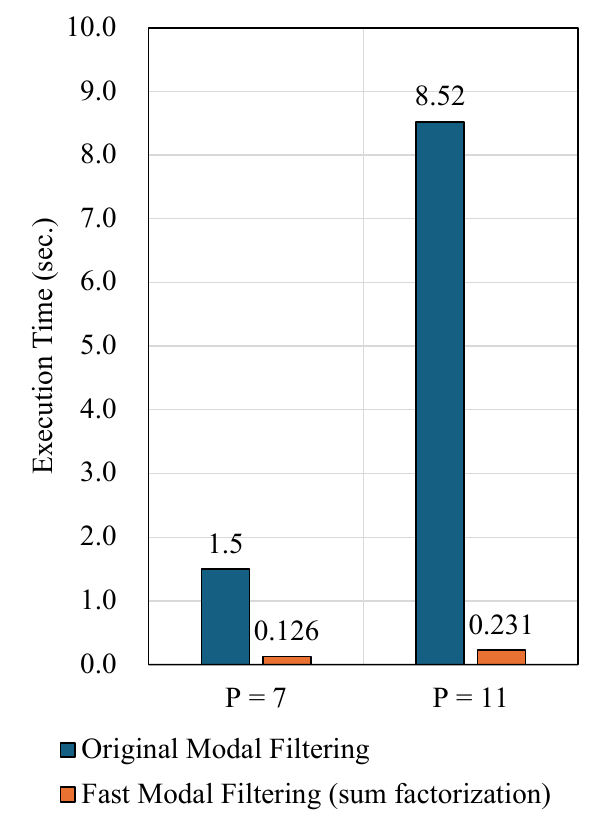}
  \caption{Execution time of the original and fast modal filtering (sum factorization implemented) at $P=7$ and $P=11$ \cite{ren_implementing_2024}.} \label{fig:time-fastmodelfilter}
\end{figure}

Fig.~\ref{fig:time-fastmodelfilter} shows its execution time for $P=7$ and $P=11$ on the Fujitsu A64FX processor. For $P=7$, the sum factorization reduces it from 1.50~s to 0.125~s, while from 8.52~s to 0.231~s for $P=11$. These correspond to speedups of $11.9\times$ and $36.8\times$, respectively.

\subsection{Performance Variation by Splitting the Loop Body}\label{subsec:perf-variation}
Although the program achieved significant speedups through the application of sum factorization, detailed profiling from previous research \cite{ren_implementing_2024} revealed that there is still room for further improvement. That study reported that the tensor $n$-mode product kernel used in sum factorization suffers from processor pipeline hazards. Listing~\ref{lst:demo-tensorprod} shows pseudo-code of the tensor $n$-mode product kernel in the $x$ direction from the sum-factorization–based implementation.

\begin{lstlisting}[float, caption={Pseudo-code of the tensor $n$-mode product kernel in the $x$ direction.}, label={lst:demo-tensorprod}]
do k=1, Np
do j=1, Np
do i=1, Np
    q_tmp(i,j,k) = mat(i,1) * q(1,j,k) +&
                   mat(i,2) * q(2,j,k) +&
                   mat(i,3) * q(3,j,k) +&
                   mat(i,4) * q(4,j,k) +&
                   mat(i,5) * q(5,j,k) +&
                   mat(i,6) * q(6,j,k) +&
                   mat(i,7) * q(7,j,k) +&
                   mat(i,8) * q(8,j,k) +&
                   ! ... ...
end do
end do
end do
\end{lstlisting}

The loop body consists of a long inner product whose number of terms is determined by the polynomial order ($P$). Each term performs a multiply-add operation, which on modern processors can be executed as a fused multiply-add (FMA) instruction. In the inner product, \texttt{q\_tmp} appears on both the left-hand and right-hand sides of each multiply-add operation. Since every new result depends on the updated value of \texttt{q\_tmp}, all instructions form a strict data dependency chain. In the processor pipeline, this dependency forces each FMA instruction to wait until the previous one has completed. The resulting stall time is equal to the instruction latency, which is the fundamental source of the pipeline hazard when the latency is long. Similar patterns occur in the other coordinate directions, leading to the same issue.

In practice, compilers will not typically generate such a naive sequence of dependent instructions. Instead, they often use multiple registers to hold intermediate results and thereby reduce dependency stalls. However, registers and reservation stations in real processors are limited resources. When the instruction sequence becomes very long, the compiler may be forced to reuse registers for accumulation, reintroducing dependency chains and pipeline hazards.

The solution is straightforward: split the inner product in the loop body into multiple parts, as illustrated in the pseudo-code in Listing~\ref{lst:demo-tensorprod-split}. This approach relaxes data dependencies and improves the efficiency of instruction-level parallelism (ILP).

\begin{lstlisting}[float,caption={Pseudo-code of the tensor $n$-mode product kernel in the $x$ direction with loop-body splitting.}, label={lst:demo-tensorprod-split}]
do k=1, Np
do j=1, Np
do i=1, Np
    tmp1 = mat(i,1) * q(1,j,k) +&
           mat(i,2) * q(2,j,k) +&
           mat(i,3) * q(3,j,k)
    tmp2 = mat(i,4) * q(4,j,k) +&
           mat(i,5) * q(5,j,k) +&
           mat(i,6) * q(6,j,k)
    tmp3 = mat(i,7) * q(7,j,k) +&
           mat(i,8) * q(8,j,k) +&
    ! ... ...
    q_tmp(i,j,k) = tmp1 + tmp2 + tmp3 + !...
end do
end do
end do
\end{lstlisting}

\begin{figure}[t!]
  \centering
  \includegraphics[width=0.8\linewidth]{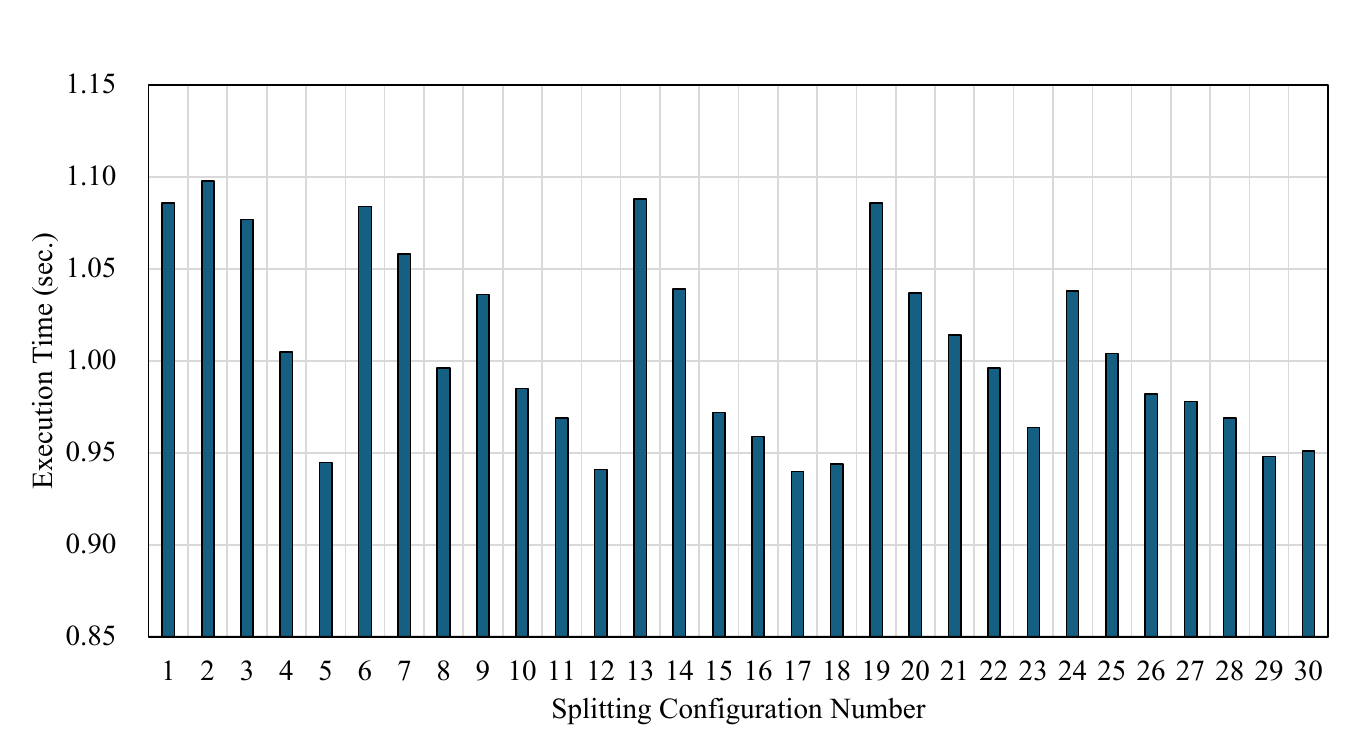}
  \caption{Execution time of the fast modal filtering (sum factorization implemented) at $P=8$ in different splitting configurations on the Fujitsu A64FX processor \cite{ren_performance_2025}.} \label{fig:time-splittingp8}
\end{figure}

Again, taking the modal filtering process in SCALE-DG as an example, Fig.~\ref{fig:time-splittingp8} shows its execution time under different splitting configurations when $P=8$ on the Fujitsu A64FX processor. Observable performance variations can be clearly recognized from the results, demonstrating that the choice of splitting configuration significantly affects performance.

The ideal splitting configuration would assign the result of each multiply-add operation to distinct variables or registers. However, as mentioned above, the number of registers in a real processor is finite, and the final accumulation step can introduce similar dependency issues if it becomes too long. Therefore, an appropriate splitting strategy must be determined to achieve the best performance. Previous research \cite{ren_performance_2025} used a trial-and-error approach to identify the optimal splitting configuration, including the number of splits and the length of each part.

Since the searching space is not large, the trial-and-error approach was enough to find out the optimal configurations. However, more systematic and predictive approach is useful for practical tunings. This motivated us to propose a model to predict the performance variations and find out the optimal configurations.

\section{Dependency Chain-Based Analytical Model}
In this section, we first introduce the concept of dependency chain and how we bridge the loop body splitting to the dependency chain. Then we introduce the proposed dependency chain-based analytical model to predict the cycles per instruction (CPI) of the fused multiply-add (FMA) instruction in the tensor $n$-mode product kernel. Finally, we verify the model with micro-benchmark on both Fujitsu A64FX and Intel Xeon Gold 6230 processors.

\subsection{Mapping Loop body Splitting to the Dependency Chain}\label{subsec:mapping-split}
\begin{figure}[t!]
  \centering
  \includegraphics[width=0.4\linewidth]{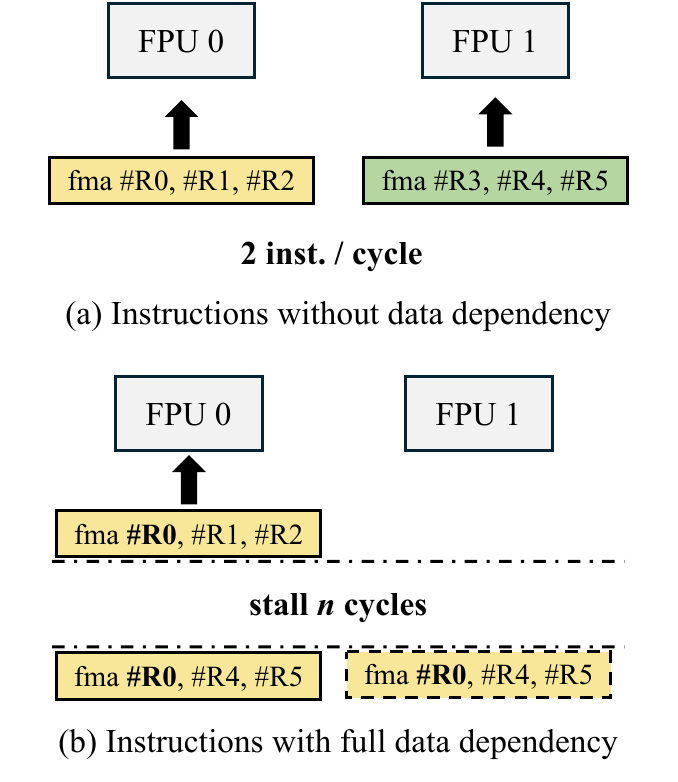}
  \caption{Illustration of floating point arithmetic pipeline status with and without data dependency.} \label{fig:demo-pipelinestall}
\end{figure}

As introduced in Section~\ref{subsec:perf-variation}, the performance degradation of the tensor $n$-mode product kernel caused by pipeline hazards can be traced to long data dependency chains in the loop body. The dependency chain, also referred as critical path (CP), is a sequence of instructions in which each depends on the results of the previous ones. The concept of it is not new. It commonly arises in processor architecture design \cite{karkhanis_automated_2007} and static code analyzers \cite{noauthor_rrze-hpcosaca_2025}. To illustrate how data dependency chains affect instruction execution time, we provide a highly simplified pipeline model of the floating-point unit (FPU) in Fig.~\ref{fig:demo-pipelinestall}.

Modern processors usually include multiple identical functional units to support parallel instruction execution. On the Fig.~\ref{fig:demo-pipelinestall}(a), all FMA instructions are independent, allowing each instruction to be dispatched freely to any FPU. In this case, two instructions can be issued and executed per cycle, yielding an average execution time of 0.5 cycles per instruction, which corresponds to the instruction throughput.

By contrast, the Fig.~\ref{fig:demo-pipelinestall}(b) depicts a fully data-dependent instruction stream. Although subsequent instructions can, in principle, be dispatched to either FPU0 or FPU1, they must wait until the previous instruction completes. The waiting time is determined by the instruction latency, which on the Fujitsu A64FX processor is 9 cycles for an FMA operation.

\begin{figure}[t!]
  \centering
  \includegraphics[width=0.4\linewidth]{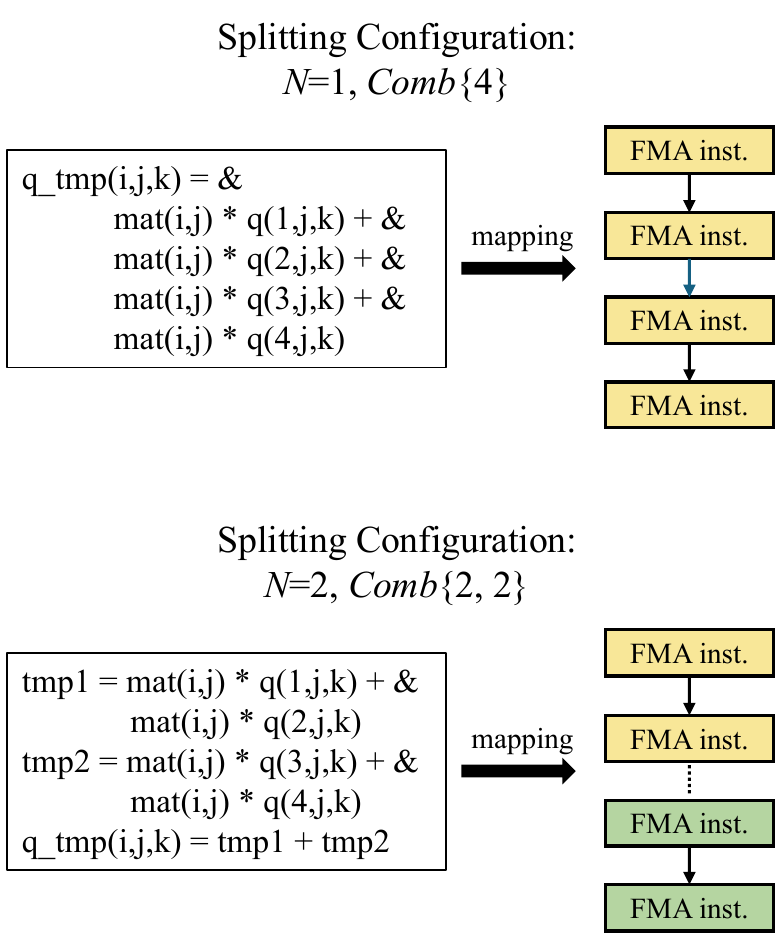}
  \caption{Illustration of mapping loop body splitting configurations to the FMA instruction dependency chains.} \label{fig:mappingfma}
\end{figure}

In the tensor $n$-mode product kernel, each term of the inner product is a multiply-add operation and corresponds to an FMA instruction. Figure~\ref{fig:mappingfma} illustrates the mapping when the loop body contains four multiply-add operations. The splitting configuration consists of two components. The first is the split count $N$, which indicates how many parts the loop body is divided into. The second is a set $Comb\{l_1, l_2, \ldots, l_n \mid 1 \leq n \leq N\}$, referred to as the split-length combination, which specifies how many terms are included in each split.

With a splitting configuration of $N=1$ and $Comb\{4\}$, no splitting is applied. In this case, the loop body corresponds to a fully data-dependent instruction stream, resulting in long stalls as introduced earlier. In contrast, with a two-way splitting configuration ($N=2$, $Comb\{2,2\}$), the original long dependency chain is transformed into several shorter chains, including the accumulation chain. Although these chains are not modeled individually, the overall effect of two-way splitting is a reduction in the longest dependency depth.

Based on these observations, we can directly bridge the loop-body splitting pattern to the corresponding dependency chains. The execution time of the kernel can then be approximated by examining the execution time of these FMA instructions.

\subsection{Model Description}\label{sec:model-desc}
These observations above allow us to construct an analytical model that predicts the execution time of FMA instructions. Fig.~\ref{fig:define-ns-ln} illustrates an instruction stream of a tensor $n$-mode product, which contains multiple data dependency chains. We denote the total number of instructions in the stream as $N_s$, and the length of the $n$th dependency chain as $l_n$ which is directly determined by the split-length combination. The execution time of each FMA instruction is assumed to be dominated by the longest dependency chain. Accordingly, Equation~\eqref{eq:ratio} defines the ratio between the length of the longest dependency chain and the total length of the instruction stream:

\begin{figure}[t!]
  \centering
  \includegraphics[width=0.5\linewidth]{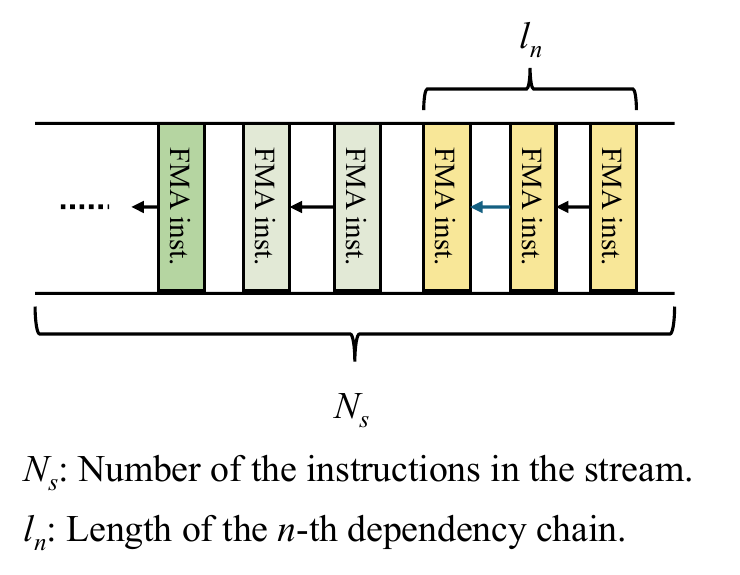}
  \caption{Definition of $N_s$ and $l_n$ in the instruction stream.} \label{fig:define-ns-ln}
\end{figure}

\begin{equation*}
Ratio = \frac{max\{l_1, l_2, \ldots,l_n\}}{N_s}
\tag{1}\label{eq:ratio}
\end{equation*}

This ratio reflects the fraction of instructions that are constrained by dependency chains, as opposed to those that can execute at the architectural throughput. A ratio close to zero corresponds to a throughput-bound case, whereas a ratio close to one corresponds to a latency-bound case. Since loop-body splitting directly shortens dependency chains, different splitting configurations are naturally represented by variations in this ratio. Only the longest dependency chain is considered, because the execution time of an instruction stream is determined by the execution time of the longest chain, as concluded by Michaud et al. \cite{michaud_exploration_2001}.

The average execution time $T_{FMA}$ of each FMA instruction can then be calculated using Equation~\eqref{eq:tfma}:

\begin{equation}
T_{FMA} = Ratio \times T_{latency} + (1-Ratio) \times T_{throughput}
\tag{2}\label{eq:tfma}
\end{equation}

Here, $T_{latency}$ represents the latency of an FMA instruction, while $T_{throughput}$ denotes the throughput of FMA instructions. Both values depend on the processor’s pipeline design and the number of FPUs.

This model is valid under the following assumptions:
\begin{itemize}
\item The data transfer timeis overlapped by computation.
\item FMA instructions dominate the instruction stream.
\end{itemize}

\subsection{Verification with Microbenchmark}
\begin{figure}[t!]
  \centering
  \includegraphics[width=0.6\linewidth]{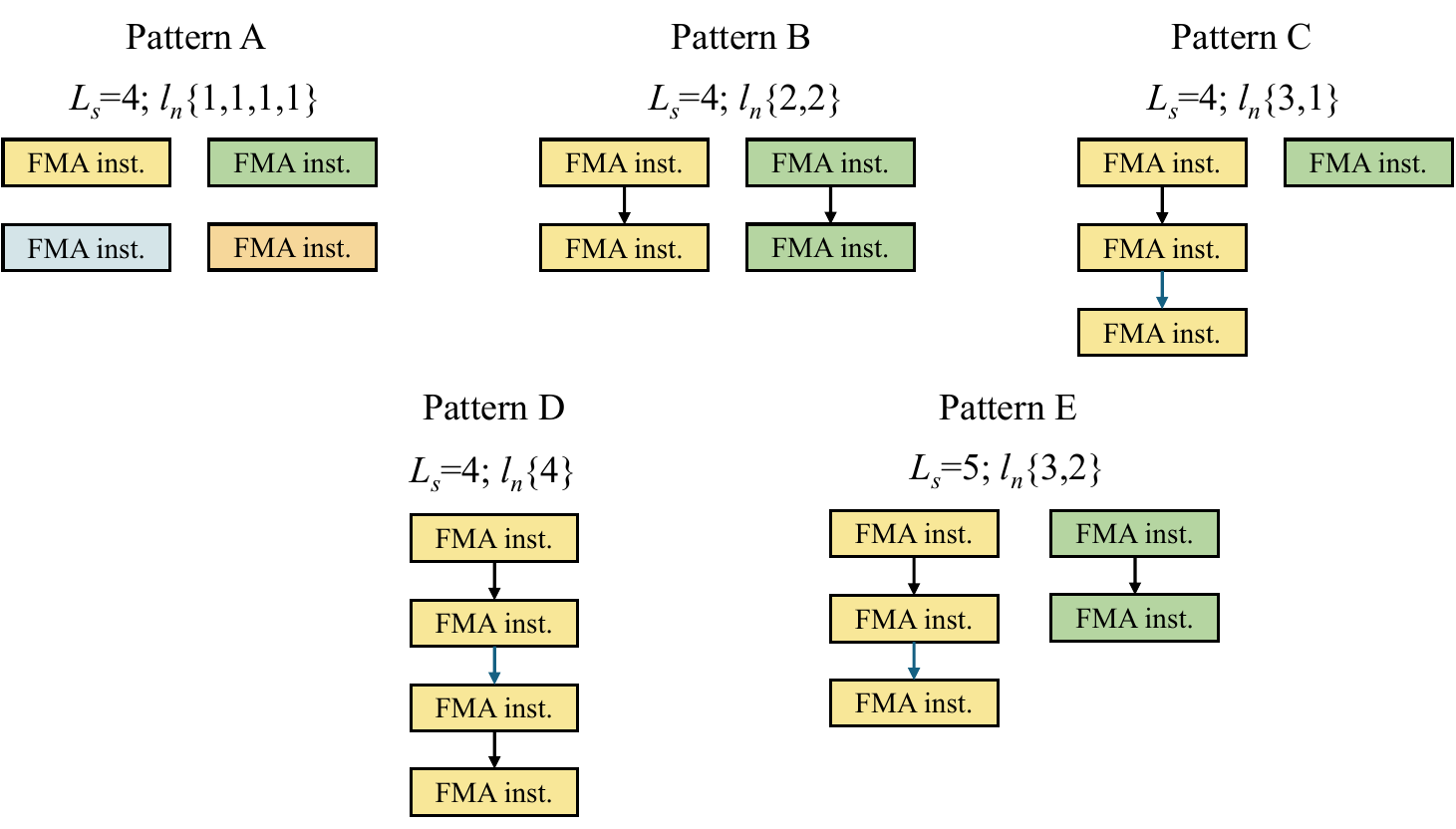}
  \caption{Dependency chain patterns used in the microbencmark.} \label{fig:microbench-patterns}
\end{figure}

A microbenchmark was developed to validate the analytical model. As shown in Fig.~\ref{fig:microbench-patterns}, five patterns are evaluated in this benchmark. Patterns A and D represent the pure throughput and full-latency conditions, respectively. Pattern B represents the case with two dependency chains of equal length. Pattern C represents a single dependency chain together with one independent instruction. Finally, Pattern E represents two dependency chains with different lengths. The values of $N_s$ and $l_n$ for each pattern are also provided in the Fig.~\ref{fig:microbench-patterns}.

Two processors are used for verification: the Fujitsu A64FX and the Intel Xeon Gold 6230. The FMA instruction throughput and latency are summarized in Table~\ref{tab:tplt}. For the A64FX, the instruction characteristics are obtained from its microarchitecture manual \cite{noauthor_a64fx_nodate}. The Intel Xeon Gold 6230, based on the Cascade Lake architecture, is characterized using data from uops.info \cite{abel_uopsinfo_2019}. Both processors feature two floating-point units, giving an instruction throughput of 0.5 cycles per FMA. However, the A64FX exhibits a higher latency of 9 cycles compared to the Intel Xeon, due to differences in pipeline design.

\begin{table}[t!]
    \caption{FMA instruction throughput and latency (in cycles) of the Fujitsu A64FX and Intel Xeon Gold 6230.}
    \label{tab:tplt}
    \centering
    \begin{tabularx}{0.5\linewidth}{l X X}
    \hline\hline
                    & $T_{throughput}$ & $T_{latency}$ \\
    \hline
    Fujitsu A64FX        & 0.5          & 9 \\
    Intel Xeon Gold 6230 & 0.5          & 4 \\
    \hline
    \end{tabularx}
    
\end{table}

To evaluate execution time at the CPU cycle level, we use information from counter timer registers for more precise measurements. On the Fujitsu A64FX, cycle counts are obtained from the \texttt{CNTVCT\_EL0} register via the \texttt{mrs} instruction. Since this register operates at its own sampling frequency, the true CPU clock cycles must be recovered by reading the frequency from the \texttt{CNTFRQ\_EL0} register. Our A64FX runs at 2.2 GHz, so the actual CPU cycles of the benchmark program can be computed using Equation~\eqref{eq:tmicrobench}:

\begin{equation*}
T_{microbench} = cntvct\_cycles \times \frac{CPU\_frequency}{cnvct\_frequency}
\tag{3}\label{eq:tmicrobench}
\end{equation*}

The execution time (in cycles) of each FMA instruction is then obtained by dividing $T_{microbench}$ by the number of instructions in the microbenchmark program. On the Intel Xeon processor, the same approach is applied, but cycle counts are obtained using the \texttt{\_\_rdtsc} intrinsic. Equation~\eqref{eq:tmicrobench} is again used to compute cycles. However, because Intel CPUs support dynamic frequency scaling (turbo boost), determining the actual CPU frequency is less straightforward and requires profiling tools such as \texttt{VTune}.

\begin{table}[t!]
\caption{Estimated and measured execution Time (in cycles) of the 5 dependency patterns on the Fujitsu A64FX and Intel Xeon Gold 6230 processors.}
\label{tab:microbench}
\centering
\begin{tabularx}{\linewidth}{X X X X X X}
\hline\hline
Pattern & $Ratio$ & \multicolumn{2}{c}{Fujitsu A64FX} & \multicolumn{2}{c}{Intel Xeon Gold 6230} \\
\cline{3-6}
        &       & Estimated $T_{FMA}$ & Measured $T_{FMA}$ & Estimated $T_{FMA}$ & Measured $T_{FMA}$ \\
\hline
A   & 0.0   & 0.50  & 0.50  & 0.50  & 0.53 \\
B   & 0.50  & 4.8   & 4.5   & 2.3   & 2.0  \\
C   & 0.75  & 6.9   & 6.8   & 3.1   & 3.0  \\
D   & 1.0   & 9.0   & 9.0   & 4.0   & 4.0  \\
E   & 0.60  & 5.60  & 5.4   & 2.6   & 2.4  \\
\hline
\end{tabularx}
\end{table}

Table~\ref{tab:microbench} shows the results of the estimated and measured execution time per FMA instruction for all five patterns. The dependency chain-based analytical model estimated very close values compared with the measured values in pattern A and D which are pure throughput and full-latency conditions, respectively, despite which processor it is. On both processors, the estimated execution time per FMA in pattern B, C, and E is slightly higher than the real measurements. It is because the $T_{throughput}$ term in the model, where in actual execution the shorter terms in pattern B, C, and E can be overlapped by the longer terms.

\subsection{Limitations of the analytical dependency chain-based model}\label{subsec:model-limit}
Although the analytical dependency-chain model explains the performance variations caused by loop-body splitting in the tensor mode-$n$ product of SCALE-DG and performs well on the microbenchmark program, it is difficult to apply it directly to real applications for the following reasons:

\begin{itemize}
\item Register usage is restricted in the benchmark program, whereas in real programs more registers are available. Moreover, in the benchmark the dependency chains are arranged in a regular order, but in practice, due to OoO execution, the actual instruction stream may appear chaotic. This makes it difficult to identify the maximum $l_n$ manually.
\item Compilers apply a variety of optimization techniques such as loop unrolling and software pipelining (SWP), which affect the final instruction scheduling. As a result, the actual length of the instruction stream ($N_s$) is uncertain.
\end{itemize}

Since both factors needed to calculate the $Ratio$ are uncertain, it is challenging to use the analytical model directly for performance prediction, especially for kernels with a large instruction base. To make the model more practical, two approaches may overcome this limitation: (1) using static code analysis tools to extract information from the kernel’s assembly code, or (2) applying machine learning to predict the $Ratio$.

\section{Learning-Augmented Performance Model}
In this section, we introduce a machine learning–based approach to overcome the limitations of the analytical model and make it more practical. First, we define a target function for machine learning, and then we describe the feature design and model structure.

\subsection{Target Function}\label{subsec:target-func}
In the original analytical model, the average execution time per FMA ($T_{FMA}$) depends on the dependency ratio as shown in Equation~\eqref{eq:ratio}. However, as discussed in Section~\ref{subsec:model-limit}, directly calculating this ratio is challenging due to compiler optimizations and OoO execution. We therefore define the learning target as the dependency ratio:

\begin{equation*}
Ratio = f(code\_features, hardware\_features)
\tag{4}\label{eq:ml-ratio}
\end{equation*}

The input features comprise both machine-level characteristics and code-level features. The calculation of $T_{FMA}$ still follows Equation~\eqref{eq:ratio}, so the new model operates in a hybrid manner. Since all ratios fall within the range $[0,1]$, they can be regarded as probability-like values. While they are not true probability distributions, this bounded shape makes them more suitable for regression compared with predicting kernel-level giga floating-point operations per second (GFLOPS) or execution time directly. 

The concept of constructing performance models based on both code and processor features dates back to the early stages of performance modeling research. The work of Noonburg et al. \cite{noonburg_theoretical_1994} proposed a theoretical model for ILP. They decomposed ILP into program parallelism, which corresponds to control and data dependencies, and machine parallelism, which corresponds to factors such as branch prediction and instruction issue policies. This framework closely resembles the principles adopted in later research, including our own approach.

Furthermore, while much research on interpretable AI (e.g., XAI \cite{ali_explainable_2023} or SHAP values \cite{lundberg_unified_2017}) focuses on extracting explanations from black-box predictors, our approach is inherently more transparent because it is grounded in an analytical model. By choosing the dependency ratio as the learning target, the machine learning model complements rather than replaces the analytical formulation, thereby preserving the interpretability of the latency/throughput decomposition. At the same time, this design makes the model more robust to compiler scheduling effects and more generalizable to other arithmetic-intensive kernels.

\subsection{Features and Model Design}
Table~\ref{tab:features} summarizes the details of the code and hardware features. The code features are intended to represent the characteristics of the tensor $n$-mode product as completely as possible. Np denotes the degrees of freedom (DoF) along one dimension, determined by the polynomial order ($P$) as $P+1$. FLOPs represents the total number of floating-point operations in the kernel, which varies with $P$. Since the loop iterations and the terms in the loop body are known before execution, this quantity is straightforward to calculate. The split count and split-length combinations, introduced in Section~\ref{subsec:mapping-split}, are key factors linking instruction dependency chains to the code implementation.

\begin{table}[t!]
    \caption{Code and hardware features for the learning target.}
    \label{tab:features}
    \centering
    \begin{tabularx}{0.6\linewidth}{|l|X|}
    \hline
    Code features & 
                        Np \newline
                        FLOPs \newline
                        Split Count: $N$ \newline
                        Split-length Combination: $Comb\{\ldots\}$ \\
    \hline
    Hardware features &
                            Frequency \newline
                            FMA latency \newline
                            FMA throughput \newline
                            FPU numbers \newline
                            Vector Length \newline
                            Register Number \newline
                            Reservation Station Size \\
    \hline
    \end{tabularx}
    
\end{table}

The hardware features include CPU frequency, the latency and throughput of FMA instructions, the number of FPUs in the CPU, and the supported vector length. The reservation station size is a more esoteric parameter, and not all processor manuals provide this information. If it is unavailable, it can be treated as a missing value in the machine learning model.

All features and target values can be represented as a row in a tabular dataset. For learning from such data, XGBoost (eXtreme Gradient Boosting) \cite{chen_xgboost_2016} is a well-suited choice. XGBoost is a tree-based machine learning framework that applies gradient boosting. We employ XGBoost instead of more complex models such as deep neural networks because our dataset size is relatively modest, and tree-based boosting methods are known to achieve high accuracy in such settings without overfitting.

Moreover, while deep learning approaches often function as opaque black boxes, XGBoost maintains interpretability through feature importance metrics, which is essential for explaining how code and hardware features influence performance. XGBoost is also efficient, because both training and inference are lightweight, making the model practical to integrate with existing performance-analysis workflows. Finally, because the regression target is the dependency ratio defined in Equation~\eqref{eq:ml-ratio}, XGBoost complements rather than replaces the analytical model, ensuring that the latency/throughput decomposition remains intact.

\section{Evaluation}
In this section, we describe how the training data were prepared and how the training procedure was conducted. We then present the training results and compare them with the Roofline and ECM models.

\subsection{Experimental Setup}\label{subsec:expr-setup}
The evaluation is conducted on the Fujitsu A64FX and Intel Xeon Gold 6230 processors. Hardware specifications are listed in Table~\ref{tab:specs}. Our Fujitsu A64FX runs in boost mode, with a frequency of 2.2 GHz. The Intel Xeon processor supports dynamic frequency scaling; however, on the Cascade Lake architecture the frequency is locked at 2.0 GHz when running full-core AVX-512 code.

\begin{table}[t!]
    \caption{Specifications of Fujitsu A64FX and Intel Xeon Gold 6230.}
    \label{tab:specs}
    \centering
    \begin{tabularx}{0.8\linewidth}{l | X X}
    \hline
    \hline
            & Fujitsu A64FX & Intel Xeon Gold 6230 \\
    \hline
    Frequency (GHz)          & 2.2 & 2.0 (all-core AVX-512) \\
    Floating Point Units     & 2   & 2 \\
    Floating Point Registers & 128 entries & 168 entries \\ 
    Reservation Station      & 40 entries & 97 entries \\
    Vector ISA extensions    & SVE & AVX-512 \\
    Vector Length (bits)     & 512   & 512 \\
    L1 Cache                 & 64 KiB & 32 KiB \\
    Cores                    & 48  & 20 \\
    NUMAs (CMGs)             & 4   & 1  \\
    \hline
    \end{tabularx}
\end{table}

To collect training data, we developed a mini-application that implements only the tensor $n$-mode product. On the A64FX platform, the Fujitsu profiling tool was used to collect performance data, while on the Intel Xeon platform, the Performance Application Programming Interface (PAPI) \cite{jagode_advancements_2025} was employed. The tensor $n$-mode product for a single element is too small to be measured accurately with these profilers. In addition, FEM computations perform the tensor $n$-mode product over the elements assigned to each CPU core. For these reasons, the mini-application handles tensor $n$-mode products executed over multiple elements. As performed in SCALE-DG, OpenMP was used to parallelize the element-level loop. One Core Memory Group (CMG) was utilized on the A64FX processor, while a single NUMA node (20 cores) was used on the Intel Xeon processor. Computations within the element are vectorized using 512-bit SVE instructions on the A64FX processor and AVX-512 instructions on the Intel Xeon processor.

For comparison with the ECM model, we applied the static code analysis tool \texttt{llvm-mca} \cite{noauthor_llvm-mca_nodate} for in-core evaluation. Although the original ECM studies typically employed OSACA \cite{noauthor_rrze-hpcosaca_2025} as the in-core analysis tool, we found that OSACA performed poorly on large instruction sequences. For example, the tensor $n$-mode product at $P=7$ produces an assembly file of 5,000–6,000 lines due to full loop unrolling by the Fujitsu compiler. On such input, OSACA ran very slowly and did not generate reasonable results. In addition, OSACA lacks support for some specific Arm SVE instructions. By contrast, although the developers of \texttt{llvm-mca} state that it is not intended for predicting execution time, we found it to perform well under our conditions.

\subsection{Training Data and Procedure}
Performance data for all split-length combinations with polynomial order $P$ ranging from 1 to 15 were collected on the Fujitsu A64FX and Intel Xeon Gold 6230, yielding approximately 1,800 rows of data.  The dependency ratio ($Ratio$) was computed retrospectively from the collected GFLOPS per core, instruction counts, processor frequencies, and instruction throughput and latency values. The detailed method is described in Appendix~\ref{apdx:a}.

It should be noted that for $P=15$ on the Intel Xeon Gold 6230, the kernel’s data size is about 70 KiB, which greatly exceeds the processor’s L1 cache capacity (32 KiB). In some cases, the calculated $Ratio$ values were greater than 1.0, which is unexpected. This may indicate that the data transfer time cannot be fully overlapped by the computation time, thereby violating the model’s assumption stated in Section~\ref{sec:model-desc}. Therefore, the data corresponding to $P=15$ on the Intel Xeon Gold 6230 were excluded from the training dataset. However, these data were still used in the results and analysis section for comparison with the ECM and Roofline models. In total, approximately 1,500 rows of data were used for training and validation.

\begin{figure}[t!]
  \centering
  \includegraphics[width=0.7\linewidth]{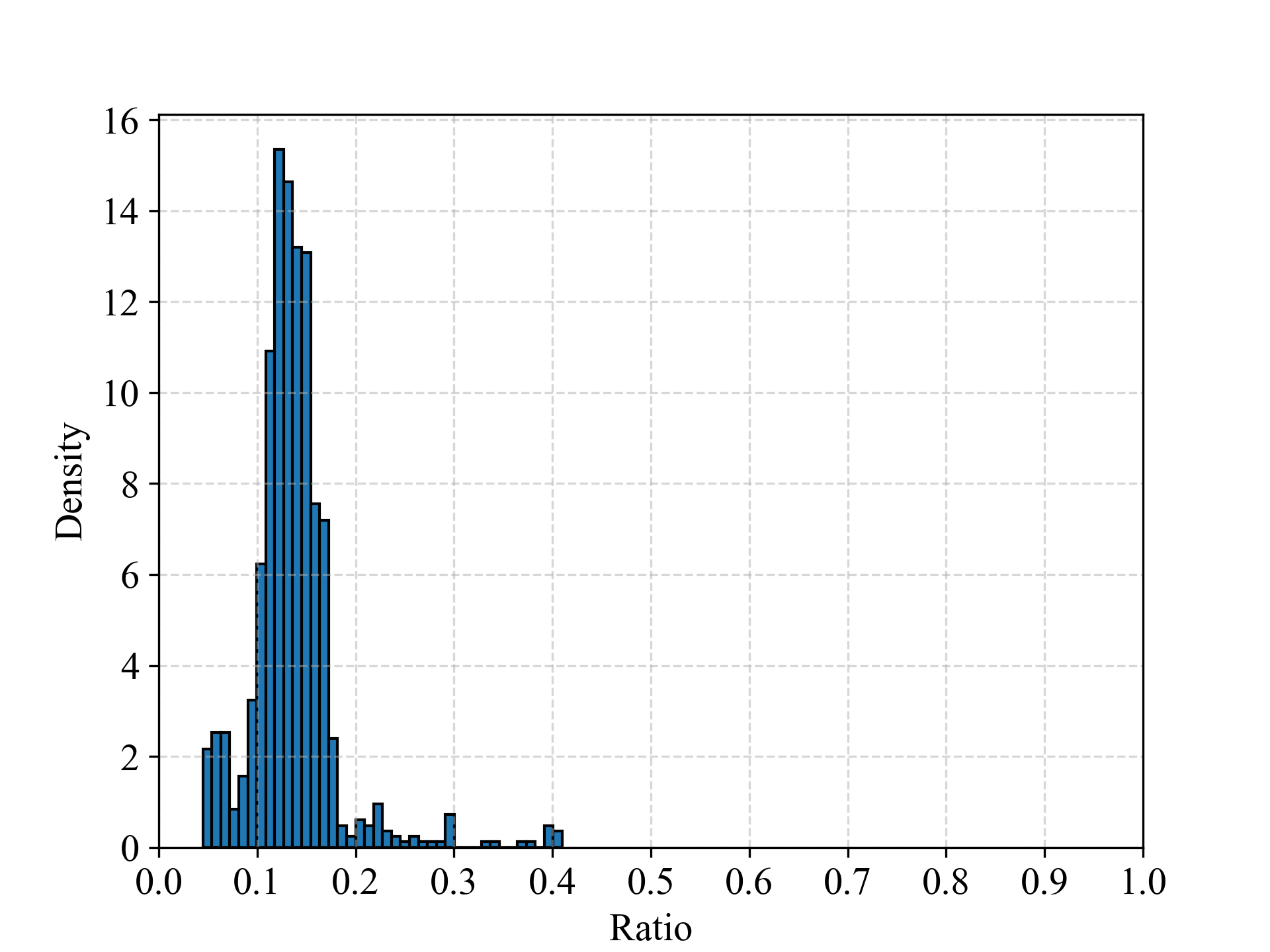}
  \caption{Distribution of $Ratio$ on the Fujitsu A64FX ($1 \leq P \leq 15$).} \label{fig:a64fx-dist-ratio}
\end{figure}

\begin{figure}[t!]
  \centering
  \includegraphics[width=0.7\linewidth]{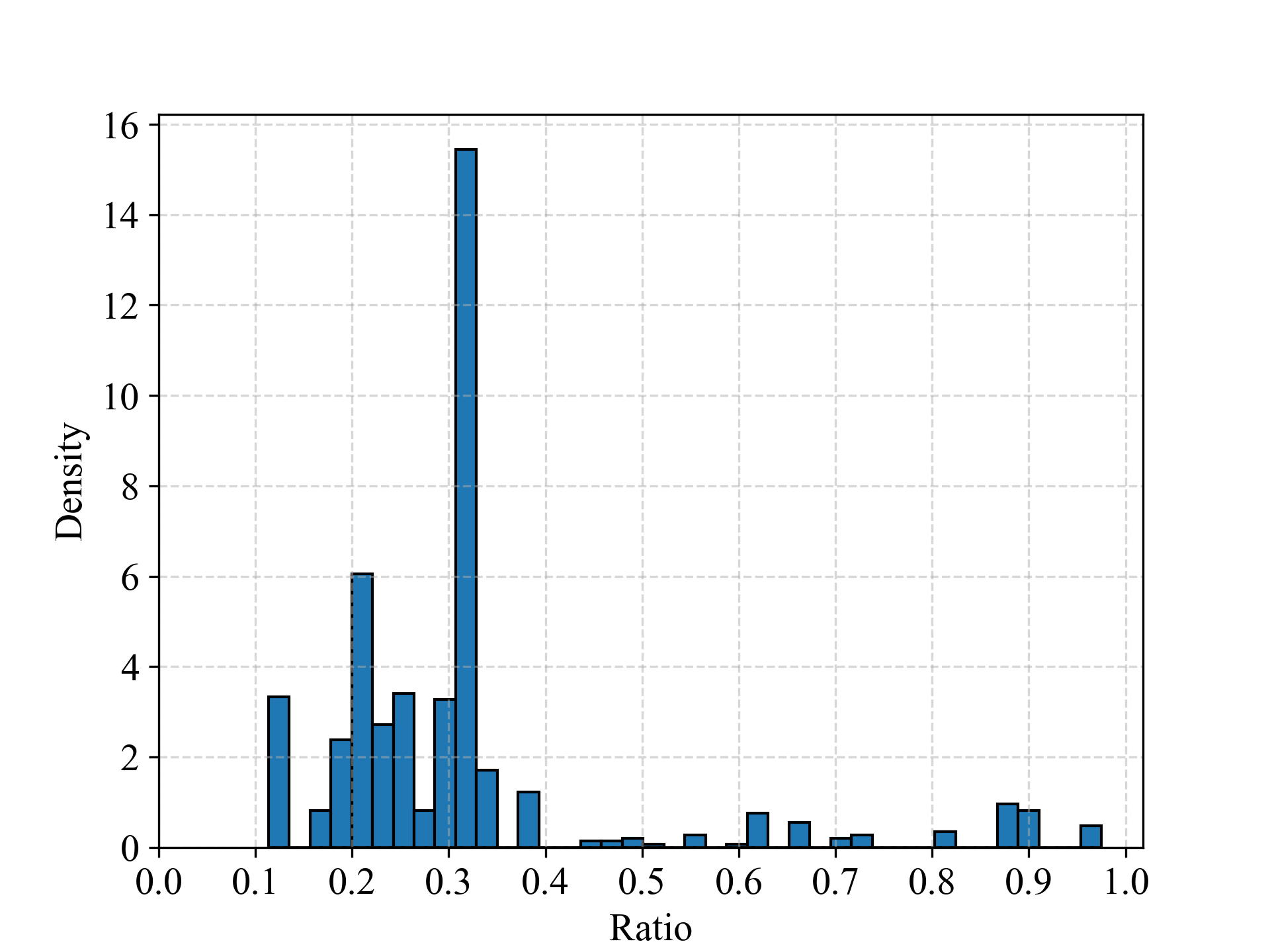}
  \caption{Distribution of $Ratio$ on the Intel Xeon Gold 6230 ($1 \leq P \leq 14$).} \label{fig:xeon-dist-ratio}
\end{figure}

Before training, we plotted the distribution of the dependency ratio ($Ratio$) to verify whether it exhibits the probability-like behavior assumed in Section~\ref{subsec:target-func}. Fig.~\ref{fig:a64fx-dist-ratio} shows the distribution on the Fujitsu A64FX. The ratios approximate a normal-like distribution, with values falling in the range $[0.05, 0.4]$ and most concentrated in $[0.1, 0.2]$. Fig.~\ref{fig:xeon-dist-ratio} shows the distribution on the Intel Xeon processor. In contrast to the A64FX, the ratio distribution on the Xeon exhibits greater variance, with three distinct peaks appearing around 0.1, 0.2, and 0.3.

For the training procedure, we used the XGBRegressor, a widely adopted regression library based on XGBoost. The dataset was split into 80\% for training and 20\% for testing. Model hyperparameters were tuned using the \texttt{RandomizedSearchCV} function in the \texttt{scikit-learn} library.

\subsection{Results and Discussion}
\begin{figure}[t!]
  \centering
  \includegraphics[width=0.6\linewidth]{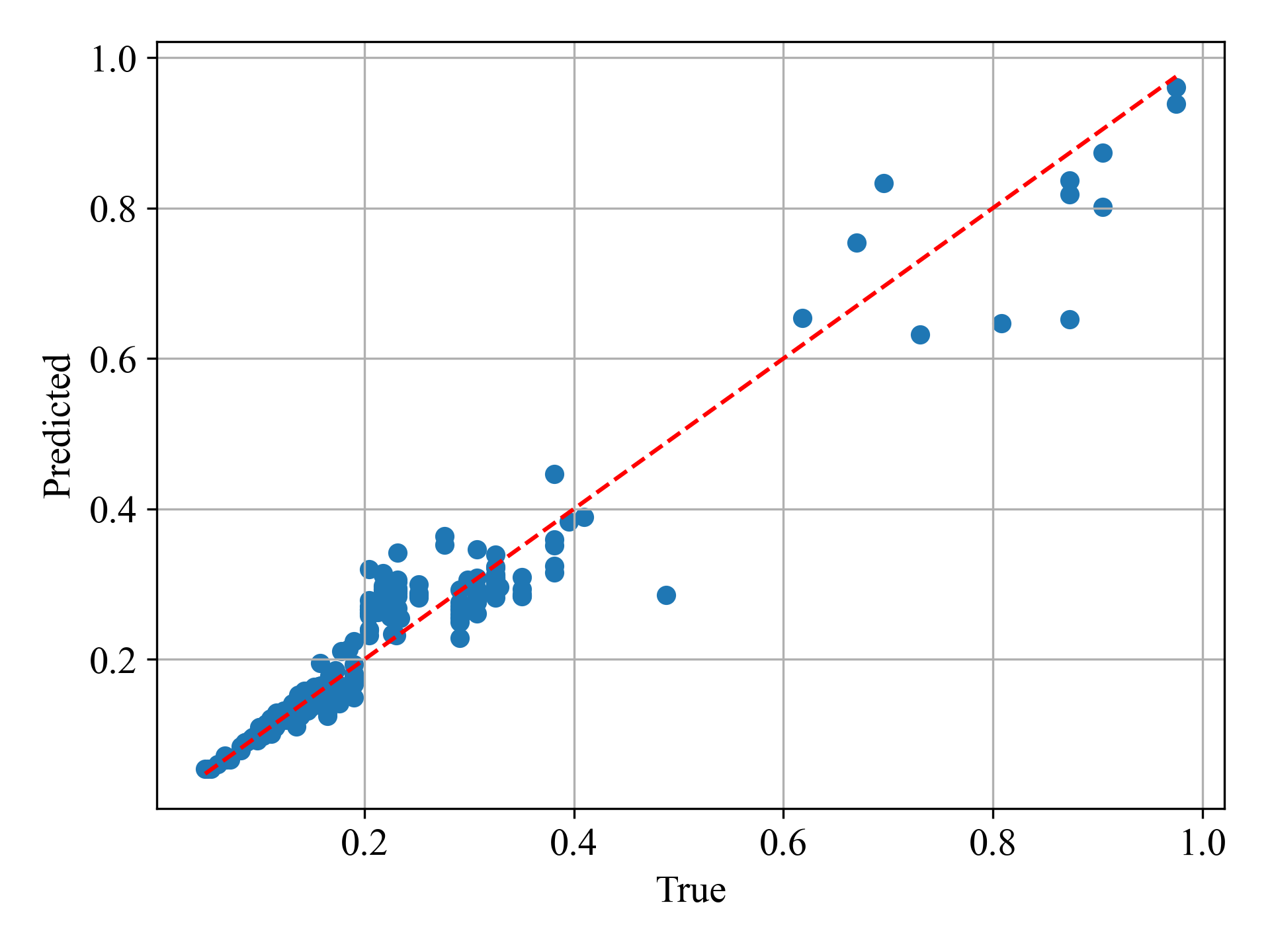}
  \caption{Training result of the learning target.} \label{fig:scatter-prediction-true}
\end{figure}

Fig.~\ref{fig:scatter-prediction-true} shows a scatter plot comparing the predicted dependency ratio with the measured ratio. The closer the points lie to the diagonal line, the better the prediction quality. As indicated by the data distributions in Fig.~\ref{fig:a64fx-dist-ratio} and Fig.~\ref{fig:xeon-dist-ratio}, most ratios fall within the range $[0.05, 0.4]$, and in this range Fig.~\ref{fig:scatter-prediction-true} shows good agreement between prediction and measurement. However, in the range $[0.6, 1.0]$, the results deviate significantly from the diagonal line, primarily due to insufficient training samples in this region.

Because the dependency ratio itself is not intuitive, we converted it into GFLOPS and compared our results with those of the widely used Roofline and ECM models. The detailed procedures for estimating GFLOPS with Roofline and ECM are provided in Appendix~\ref{apdx:b}. Fig.~\ref{fig:gflopscomp-a64fx-p7} shows the GFLOPS estimated by Roofline, ECM, and our learning-augmented model for test cases on the A64FX at polynomial order $P=7$. The lateral axis represents different split-length combinations. The results show that both the Roofline and ECM models overestimate kernel performance, while our model produces values much closer to the measurements.

\begin{figure}[t!]
  \centering
  \includegraphics[width=0.7\linewidth]{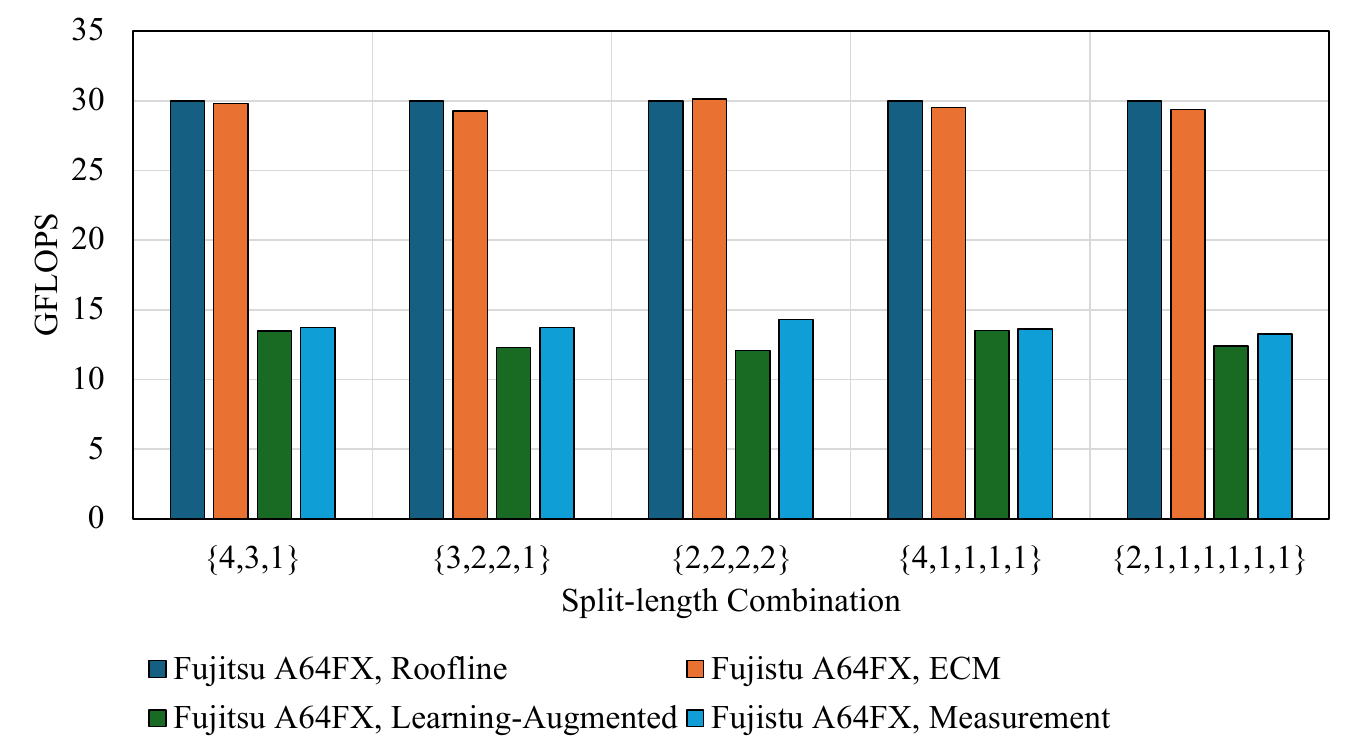}
  \caption{GFLOPS per core at $P=7$ estimated by the Roofline, ECM, and Learning-Augmented models comapred to the measured values on the Fujitsu A64FX.} \label{fig:gflopscomp-a64fx-p7}
\end{figure}

\begin{figure}[t!]
  \centering
  \includegraphics[width=0.7\linewidth]{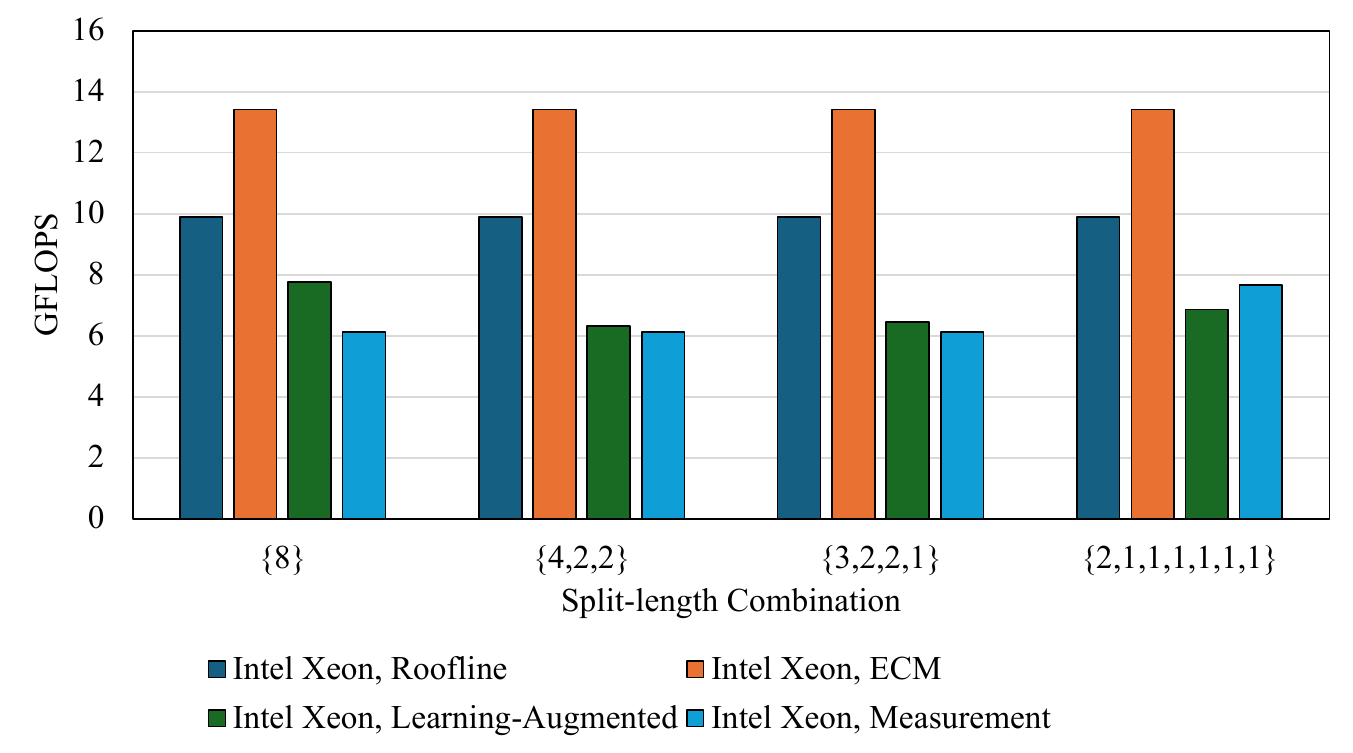}
  \caption{GFLOPS per core at $P=7$ estimated by the Roofline, ECM, and Learning-Augmented models comapred to the measured values on the Intel Xeon Gold 6230.} \label{fig:gflopscomp-xeon-p7}
\end{figure}

Fig.~\ref{fig:gflopscomp-xeon-p7} illustrates the same comparison on the Intel Xeon processor. In this case, the ECM model consistently predicts higher GFLOPS than the Roofline model. This is because the Roofline model bases its prediction solely on memory bandwidth, whereas the ECM model also accounts for data transfers at each cache level. Since in our experiments almost all hot data reside in the L1 cache, whose bandwidth is substantially larger than main memory, the ECM model reports higher GFLOPS than Roofline. Nevertheless, neither Roofline nor ECM closely matches the real measurements, whereas the learning-augmented model does.

\begin{figure}[t!]
  \centering
  \includegraphics[width=0.7\linewidth]{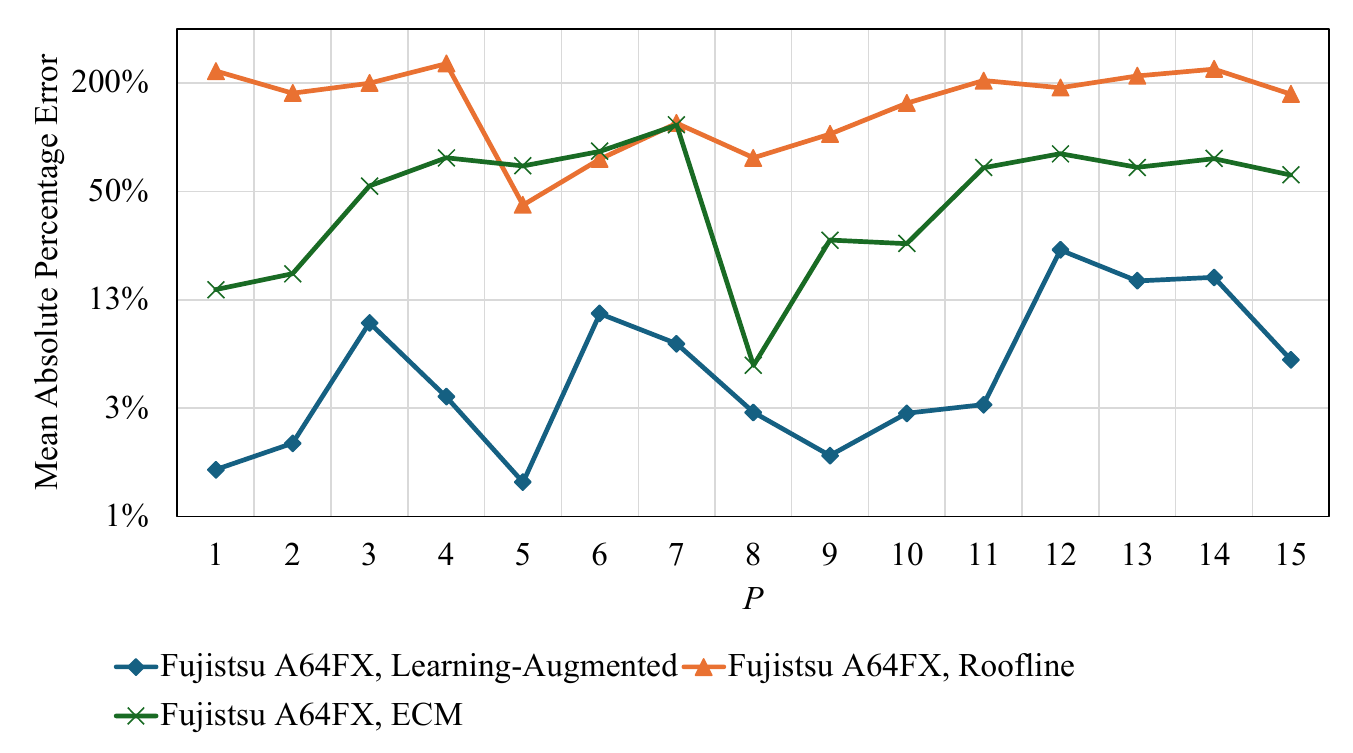}
  \caption{Comparison of Learning-Augmented, Roofline, and ECM model across different $P$s and split-length combinations on the Fujitsu A64FX.} \label{fig:gflopsmape-a64fx}
\end{figure}

\begin{figure}[t!]
  \centering
  \includegraphics[width=0.7\linewidth]{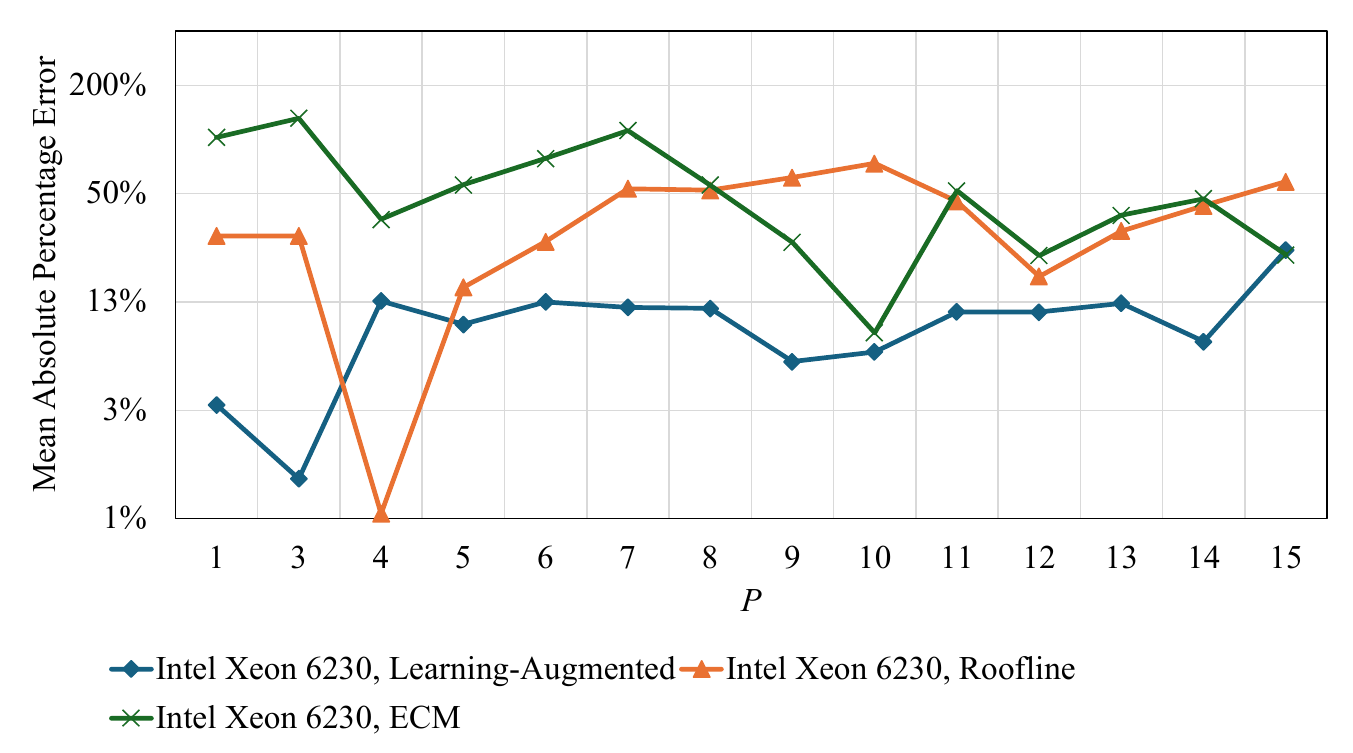}
  \caption{Comparison of Learning-Augmented, Roofline, and ECM model across different $P$s and split-length combinations on the Intel Xeon Gold 6230.} \label{fig:gflopsmape-xeon}
\end{figure}

To compare the learning-augmented model with the ECM and Roofline models across different polynomial orders ($P$) and splitting configurations, Fig.~\ref{fig:gflopsmape-a64fx} and Fig.~\ref{fig:gflopsmape-xeon} are provided. Fig.~\ref{fig:gflopsmape-a64fx} shows the mean absolute percentage error (MAPE) of GFLOPS predicted by the learning-augmented, Roofline, and ECM models compared with the measured values, grouped by $P$ on the Fujitsu A64FX processor. The data are from the 20\% testing subset, the vertical axis is presented in logarithmic scale. The learning-augmented model demonstrates excellent prediction accuracy across all $P$ values, with errors ranging from 1\% to 24\%. The Roofline model exhibits larger errors but shows reasonable agreement at $P=5$. The ECM model performs better than the Roofline model, providing very accurate predictions for $P=1,2$ and $P=8,9,10$. This occurs because, for $P=1,2$, the data transfer volume is too small for memory effects to dominate, so the in-core execution time estimated by \texttt{llvm-mca} governs the total performance. For $P=8,9,10$, the data size causes the data transfer time to become comparable to the in-core execution time. Overall, however, the learning-augmented model outperforms the ECM model in predictive accuracy.

Fig.~\ref{fig:gflopsmape-xeon} shows the same MAPE comparison, this time on the Intel Xeon Gold 6230 processor. The learning-augmented model again demonstrates excellent overall performance for $P=1$ to $P=14$, with prediction errors ranging from 1\% to 13\%. Interestingly, the Roofline model exhibits smaller errors than the ECM model in many cases, opposite to the trend observed on the A64FX, and even provides highly accurate predictions at $P=4, 5$. The ECM model shows larger errors than the Roofline model in most cases, primarily because its estimated performance is based on cache bandwidth which has been stated when describing results in the Fig.~\ref{fig:gflopscomp-xeon-p7}. As a result, the Roofline predictions are typically lower than those of the ECM model. Nevertheless, the ECM model provides very good predictions at $P=9,10,$ and $12$.

A particularly noteworthy case in Fig.~\ref{fig:gflopsmape-xeon} is at $P=15$, where the data violate the assumptions of our learning-augmented model. The MAPE of the learning-augmented model is slightly above 20\%, representing a noticeable degradation compared with predictions at other $P$ values. In this regime, data transfers between memory hierarchies can no longer be ignored, and this is precisely the situation that the ECM model is designed to handle. Although the MAPE is around 20\%, a direct comparison of absolute GFLOPS provides additional insight: the measured performance is approximately 12–15 GFLOPS, while the ECM model predicts about 15 GFLOPS and the learning-augmented model predicts 10–11 GFLOPS. Consequently, in this case, the ECM model demonstrates superior accuracy.

\subsection{General Application Performance Modeling}
For general application performance evaluation, widely used models such as the standard Roofline model \cite{williams_roofline_2009} and the Cache-Aware Roofline model \cite{ilic_cache-aware_2014} primarily focus on memory and cache bandwidth. The Roofline model has been extended by Cabezas et al. \cite{cabezas_extending_2014} to incorporate microarchitectural constraints, but dependency-chain or critical-path analysis is usually handled by static code analyzers such as llvm-mca \cite{noauthor_llvm-mca_nodate}, which performs cycle-by-cycle simulations of the processor pipeline, and tools such as OSACA \cite{laukemann_automated_2018, laukemann_automatic_2019} and Facile \cite{abel_facile_2023}, which construct analytical throughput models based on static instruction analysis. Although effective, these tools require detailed assembly-level information, which limits their applicability for high-level kernel modeling.

The Execution-Cache-Memory (ECM) model \cite{hofmann_execution-cache-memory_2015} extends the Roofline framework by integrating in-core execution with data transfer modeling across memory hierarchy levels. However, its in-core predictions still rely on throughput estimations from the aforementioned tools. More recent research has begun to incorporate instruction throughput more explicitly. For example, Leinhauser et al. \cite{leinhauser_metrics_2021} proposed an Instruction Roofline model for AMD GPUs, and Ding et al. \cite{ding_instruction_2022} introduced a similar model for NVIDIA GPUs. Shi et al. \cite{nagel_arm_2026} examined the performance impact of Arm’s Scalable Vector Extension (SVE) instructions on HPC processors and proposed a machine learning–based model to classify performance bottlenecks.

\subsection{Tensor Operation Performance Tuning}
In the more specific domain of tensor operation performance tuning, memory bandwidth remains the dominant evaluation factor. Swirydowicz et al. \cite{swirydowicz_acceleration_2019} presented a Roofline-like model to evaluate tensor kernels on GPU, incorporating both global and shared memory bandwidth across different polynomial orders. Vectorization efficiency was studied by Kempf et al. \cite{kempf_automatic_2021}, who developed an automatic framework for generating and tuning tensor operation code in discontinuous Galerkin methods. Their model used a logarithmic function to represent ILP efficiency.

Learning-based performance tuning has also gained attention. Chen et al. \cite{chen_learning_2018} proposed domain-specific statistical cost models to guide tensor operation optimization in deep learning workloads, using features extracted from low-level abstract syntax trees (ASTs). Kaufman et al. \cite{kaufman_learned_2021} introduced a graph neural network (GNN)–based model for tensor program tuning on Tensor Processing Units (TPUs), outperforming conventional methods for tile-size selection and operator fusion. Similarly, Zhai et al. \cite{zhai_tlp_2023} developed a deep learning–based cost model that extracts features from schedule primitives in TVM \cite{chen_tvm_2018}. More recently, Sudusinghe et al. \cite{sudusinghe_cognate_2025} proposed a transfer learning–based cost model for sparse tensor operation tuning, which reduces the need for large training datasets. Although these learning-based works do not explicitly model instruction throughput, the code features they use likely encode instruction-level efficiency implicitly.

\subsection{Positioning of This Work}
Our work shares similarities with learning-based performance models in that it takes both code and hardware features as input and employs XGBoost as the regression model. However, it differs from prior work in several key aspects:
\begin{itemize}
\item Unlike purely analytical or purely learning-based approaches, our learning-augmented model adopts a hybrid design to predict the cycles per FMA instruction (CPI). The analytical component ensures interpretability, while the machine learning component enhances robustness and practical applicability.
\item While common methods to improve OoO execution include loop unrolling and loop reordering, our work focuses on loop-body splitting, a less explored optimization technique. The proposed model explicitly incorporates split counts and combinations as input features.
\item Most ML-based tensor performance tuning studies originate from deep learning frameworks. In contrast, our work focuses on tensor $n$-mode product in high-order FEM, representing a different application domain.
\end{itemize}

\section{Conclusion}
In this study, we proposed a dependency chain–based analytical model to explain the performance variation observed during the optimization of tensor product factorization in high-order FEM, specifically the effect of splitting the loop body. The analytical model bridges the relationship between split counts, split-length combinations, and the data dependencies of FMA instructions in tensor $n$-mode product implementations. We then augmented the analytical model using XGBoost, creating a learning-augmented approach that enhances robustness and predictive capability.

Evaluation was conducted on the Fujitsu A64FX and Intel Xeon Gold 6230 processors for polynomial orders ($P$) ranging from 1 to 15. On the A64FX processor, the learning-augmented model achieved a mean absolute percentage error (MAPE) between 1\% and 24\%, compared with 5\%–117\% for the ECM model and 42\%–256\% for the Roofline model. On the Intel Xeon processor, the learning-augmented model achieved MAPE values from 1\% to 13\% for $P=1$–14, increasing to 24\% at $P=15$ due to violation of the model’s L1-cache-residency assumption. The ECM model’s errors ranged from 8\% to 112\%, while the Roofline model ranged from 1\% to 73\%. Overall, the learning-augmented model demonstrated consistently superior predictive accuracy across both architectures and most polynomial orders.

This work provides a more systematic approach for tensor product factorization performance tuning in high-order FEM and offers potential applicability to other element–based atmospheric simulation codes, such as SCALE-DG.

In future work, we plan to expand the dataset by collecting data from a wider variety of processors and compilers. Moreover, additional instruction types beyond FMA will be integrated into the model, and cache-related effects, such as miss rates and associated latencies, will be incorporated through predictive modeling.

\section*{Acknowledgment}
This work was financially supported by JST SPRING, Grant Number JPMJSP2125. This work was also supported by the Joint Usage/Research Center for Interdisciplinary Large-scale Information Infrastructures (JHPCN) and the High-Performance Computing Infrastructure (HPCI) under project number jh250015 and jh250018. In addition, this research was funded by JSPS KAKENHI Grants JP23K11126 and JP24K02945. The author (Initial) would like to take this opportunity to thank the “THERS Make New Standards Program for the Next Generation Researchers”.

\section*{Appendices}
\appendix
\section{Computing dependency ratio retrospectively}\label{apdx:a}
Although the profiling tools cannot obtain the dependency ratio directly, we can compute it retrospectively. First, the total execution time in cycles of the kernel is derived using Equation~\eqref{eq:t-kernel}:
\begin{equation}
T_{kernel} = \frac{FLOPs \times CPU\_frequency}{GFLOPS}
\tag{A.1}\label{eq:t-kernel}
\end{equation}
where FLOPs refers to the total floating-point operations of the kernel, GFLOPS refers to the giga floating-point operations per second that obtained from the profilers.

In our model, it is assumed that nearly all execution time is dominated by FMA instructions. Therefore, $T_{kernel}$ can be approximated as the total execution time of all FMA instructions. If the number of FMA instructions is known, the average execution time (in cycles) per FMA can be calculated using Equation~\eqref{eq:real-tfma}, where $N_{FMA}$ denotes the FMA instruction count:
\begin{equation}
T_{FMA} = \frac{T_{kernel}}{N_{FMA}}
\tag{A.2}\label{eq:real-tfma}
\end{equation}
In this work, if the kernel is fully unrolled, $N_{FMA}$ is obtained by counting the number of FMA instructions in the assembly files. Otherwise, the $N_{FMA}$ is calculated by the the number of FMA instructions in the assembly files times the loop iterations.

Finally, since the instruction throughput and latency of the processor are given, the dependency ratio can be computed using Equation~\eqref{eq:real-ratio}:
\begin{equation*}
Ratio = \frac{T_{FMA} - T_{throughput}}{T_{latency} - T_{throughput}}
\tag{A.3}\label{eq:real-ratio}
\end{equation*}

\section{Estimating GFLOPS using Roofline and ECM model}\label{apdx:b}
\subsection{Estimating with Standard Roofline model}
In this work, the standard Roofline model \cite{williams_roofline_2009} is employed to estimate the GFLOPS of the tensor $n$-mode product. Arithmetic intensity ($AI$), also called FLOPs per byte, is the only factor required in this model. We consider only the data read and written at the kernel level. The tensor $n$-mode product reads the hexahedral element data \texttt{q\_in} and \texttt{q\_tmp}, each of size $(P+1)^3$, and then writes back the updated element data \texttt{q\_in} and \texttt{q\_tmp}. The three operator matrices, each of size $(P+1)^2$, are relatively small compared with the element data and are loaded only once during the first execution. These matrices are assumed to remain resident in the cache, so their data transfer is ignored. All data are stored in double precision (8 bytes per entry).

The $AI$ can be computed as follows:
\begin{equation}
AI = \frac{FLOPs}{(P+1)^3 \times 4 \times8}
\tag{B.1}\label{eq:ai}
\end{equation}
The estimated GFLOPS can then be obtained by multiplying AI with the available memory bandwidth. Finally the standard Roofline model can be represented as:
\begin{equation}
GFLOPS \leq min \begin{cases}
Peak & GFLOPS \\
Peak & GB/s \times AI
\end{cases}
\tag{B.2}
\end{equation}

\subsection{Estimating with ECM Model}
The ECM model is more complex. It consists of two main components: the execution time that can be overlapped (e.g., in-core arithmetic) and the execution time that cannot be overlapped, including data transfer times. Following the work of C. Alappat et al. \cite{alappat_executioncachememory_2022}, the basic form of the ECM model on the Fujitsu A64FX is given in Equation~\eqref{eq:basic-ecm}:
\begin{equation*}
T_{ECM} = max\{T_{c\_OL}, f(T_{L1\_LD}, T_{L1\_ST}, T_{L2}, T_{Mem})\}
\tag{B.3}\label{eq:basic-ecm}
\end{equation*}
Here, $T_{c\_OL}$ mainly refers to floating-point operations that can overlap with data transfers. The function $f(T_{L1\_LD}, T_{L1\_ST}, T_{L2}, T_{Mem})$ represents the data transfer time, which can behave differently depending on the processor. Transfers between different memory hierarchies may or may not overlap, depending on the overlap hypothesis of the processor. For the A64FX, the overlap hypothesis follows Equation~\eqref{eq:olhypo-a64fx} which also comes from the work \cite{alappat_executioncachememory_2022}. For the Intel Xeon Gold 6230 (Cascade Lake), we did not find a precise description of the hypothesis in the literature, but we identified an equation from the official GitHub repository \cite{noauthor_rrze-hpcecm-model_nodate} of ECM model. We slightly modified it by eliminating the \texttt{L1\_shared\_bw} term, because in our kernel reads dominate writes, especially at high polynomial orders. In this paper, its overlap hypothesis is expressed by Equation~\eqref{eq:olhypo-xeongold}. 
\begin{equation*}
\begin{split}
f(T_{L1\_LD}, T_{L1\_ST}, T_{L2}, T_{Mem}) = \\
      max\{T_{L1\_LD} + max\{ T_{L1\_ST}, T_{L2}\}, T_{Mem}\}
\end{split}
\tag{B.4}\label{eq:olhypo-a64fx}
\end{equation*}

\begin{equation}
\begin{split}
f(T_{L1\_LD}, T_{L1\_ST}, T_{L2}, T_{L3\_RD}, T_{L3\_WR}, T_{Mem}) = \\
      max\{T_{L1\_LD}, T_{L1\_ST}\} + T_{L2} + \\
      max\{T_{L3\_RD}, T_{L3\_WR}\} + T_{Mem}  
\end{split}
\tag{B.5}\label{eq:olhypo-xeongold}
\end{equation}

As introduced in Section~\ref{subsec:expr-setup}, $T_{c\_OL}$, $T_{L1\_LD}$, and $T_{L1\_ST}$ are obtained from the static analysis tool \texttt{llvm-mca}. Other execution times are computed by dividing the data transfer volume by the corresponding bandwidth. The estimated single-core GFLOPS is then calculated using Equation~\eqref{eq:recvr-gflops}:
\begin{equation}
GFLOPS = \frac{FLOPs}{T_{ECM}} \times CPU\_frequency
\tag{B.6}\label{eq:recvr-gflops}
\end{equation}

\bibliographystyle{unsrtnat}
\bibliography{perfmodel}  

\end{document}